\documentstyle[12pt]{article}
\begin{document}

\title{Test of Comoving Coordinate Frame by Low Magnitude-Redshift Relation}
\author{Amir H. Abbassi$^{*}$ and Sh. Khosravi \\
{\small {\it Department of Physics, School of Sciences }}\\
[0mm] {\small {\it Tarbiat Modarres University, P.O.Box 14155-4838 }}\\
[0mm] {\small {\it Tehran, I.R. Iran}}\\
$*$E-mail : ahabbasi@modares.ac.ir}
\date{}
\maketitle

\begin{abstract}
Observational evidence from a variety of sources points at an accelerating
universe which is approximately spatially flat and with $(\Omega_M , 
\Omega_\Lambda)_0 \approx (0.3 , 0.7)$. We have shown that for low 
redshifts, $z\leq 0.2$, the metric of this cosmological model is equivalent
to the de Sitter metric within one percent error. 
 Among various coordinate descriptions for the flat de Sitter model two are most widely used, one yields a non-static and RW type, while the other gives a
static and Schwarzschild type metric.We have obtained the magnitude-redshift
relation in the second coordinate frame. Our result indicates a slope of
2.5. This is in disagreement with observation which is close to the slope of
5. This test discards the second and confirms the first coordinate frame as
a comoving frame for matter in de Sitter space. We anticipate this
test disqualifies any other solution of the field equation which
asymptotically approaches to the static de Sitter metric, e.g. the
Schwarzschild-de Sitter metric for space-time around a point mass $m$.

\bigskip\ 


\end{abstract}

\pagebreak

As was recently shown by independent groups distant type
Ia supernovae provide a striking evidence for an acceleration of the universal
expansion quite contrary to the conventional expectations[1-4].A positive
cosmological constant is inferred from these measurements[5].

Up to now a satisfactory explanation has not been presented for the
existence of a non-zero cosmological constant by the particle physics.
Actually, calculations of vacuum energy, which
is directly related to the value of cosmological constant, by field theoretical methods give results
with a difference of 120 orders of magnitude from the observed values[6]. For these reasons a
cosmological constant factor has not been taken seriously in the common
discussions of cosmology, but now with these new evidences it seems
plausible to consider this factor not to be negligible.

The line element according to FRW model for the de Sitter space, which is
appropriate for $\Lambda$-dominated worlds, is 
\begin{equation}
ds^{2}=dt^{2}-R^{2}(t) [dr^{2}+r^{2}(d \theta^{2}+ \sin ^{2} \theta d \phi
^{2})]
\end{equation}
with the scale factor as 
\begin{equation}
R(t)= \exp [( \frac{\Lambda}{3} )^{1/2}t]
\end{equation}
where $\Lambda$ is the cosmological constant.

Today the observatinally favoured universe is near  $(\Omega_M , 
\Omega_\Lambda)_0 \approx (0.3 , 0.7)$ and $k=0$. The Friedmann
equation for such a cosmological model may be easily solved and 
the scalar factor in the FRW metric is given by 
\begin{equation}
\frac{R(t)}{R_0}=[\sqrt{\frac{\Omega_{M_0}}{\Omega_{\Lambda_0}}}
\sinh{\frac{\sqrt{3\Lambda}t}2}]^{2/3}
\end{equation}
where $R_0$ is the scalar factor at our epoch. For $z=0$ and $z=0.2$
which  correspond  to $R(t_0)=R_0$ and $R(t_1)=\frac{R_0}{1.2}$
respectively, equ.(3) gives $\frac{\sqrt{3\Lambda}}{2}t_0=1.21$
and $\frac{\sqrt{3\Lambda}}{2}t_1=1.99$. It can be  shown that
\begin{equation}
\frac{R(t_0)}{R(t_1)}=\{\exp{(\sqrt{\frac{\Lambda}3}(t_0-t_1))}\}
\{1+0.01\}
\end{equation}
Thus in a good approximation we may assume that in this cosmological 
model the metric for $z\leq 0.2$ is of the de Sitter form. For the rest we
restrict our discussion to this range.

For a FRW model, independent of the form of $R(t)$, there is a relation
between the distance of a light source and its redshift due to the
expansion. This relation for luminosity distance versus redshift expanded as
a power series is written as follows, 
\begin{equation}
d_L=H_0^{-1}[z+\frac 12(1-q_0)z^2+\ldots ]\,,
\end{equation}
where z is the redshift and $H_0$ and $q_0$ are the present values of
Hubble's constant and decelaration parameter. Obviously this relation is
linear in the first approximation which gives rise to Hubble's law.
Also, it can be written in terms of distance modulus, 
\begin{equation}
m-M=25-5\log H_0+5\log z+1.086(1-q_0)z+\ldots 
\end{equation}
and 
\begin{equation}
d_L=10^{1+(m-M)/5}
\end{equation}
$m$ and $M$ being apparent and absolute magnitudes of the source,
respectively.

On the other hand, there exists an isometry of de Sitter solution that gives a metric of
schawrzschild type for which the line element is, 
\begin{equation}
ds^{2}=(1-\frac{\Lambda}{3}\rho^{2})dT^{2}-(1-\frac{\Lambda}{3}%
\rho^{2})^{-1}d\rho^{2}-\rho^{2}(d \theta^{2}+ \sin ^{2} \theta d \phi ^{2})
\end{equation}
This solution is static and can be obtained from the ordinary solution by
the following transformation, 
\begin{equation}
\rho=R(t)r
\end{equation}
\begin{equation}
T=t-\sqrt{\frac{3}{4\Lambda}}\ln \mid(1-\frac{\Lambda}{3}\rho^{2})\mid
\end{equation}
The comoving frame is uniquely defined frame of reference in which a given extended set of bodies is at rest or is as much at rest as possible in the sense that their kinetic energy is small as possible. In a cosmological model which satisfies the cosmological principle, a comoving coordinate frame is defined so that the matter distribution has fixed space coordinates. Our aim is to check whether the static de Sitter coordinate frame is eligible to serve as a comoving reference frame. For this purpose by considering the static metric (8) we are going to calculate the redshift - magnitude relation again.

If we suppose that a source is at a point with distance $\rho $ from origin
and sends a signal to an observer at the origin, we can find that the
gravitational redshift is proportional to the zero-component of metric at $%
\rho $ [7]. In this case we have 
\begin{equation}
g_{00}=(1-\frac \Lambda 3\rho ^2)
\end{equation}
so that
\begin{equation}
(1+z)^2=\frac 1{1-\frac \Lambda 3\rho ^2}
\end{equation}
and therefore 
\begin{equation}
\rho =\sqrt{\frac 3\Lambda \frac{z(2+z)}{(1+z)^2}}
\end{equation}
On the other hand when a source in the distance $\rho $ emits photons, the
energy flux per unit area is diminished by the factor $(1+z)^2$ . So for the
static metric stated here , we can write the energy flux as follows 
\begin{equation}
f=\frac{{\cal L}}{4\pi \rho ^2(1+z)^2}
\end{equation}
where ${\cal L}$ is the luminosity of source. From (13) and (14) we have 
\begin{equation}
f=\frac \Lambda {12\pi }\frac{{\cal L}}{z(2+z)}
\end{equation}
In addition , the following relations are used as the definition of relative
and absolute magnitudes 
\begin{equation}
m=-2.5\log f+k_1
\end{equation}
and 
\begin{equation}
M=-2.5\log {\cal L}+k_2
\end{equation}
in which $k_1$ and $k_2$ are constants. Now we can write according to (15):
\begin{equation}
m-M=2.5\log [z(2+z)]-2.5\log (\frac \Lambda {12\pi })
\end{equation}
By expanding the function in the brackets in (18) we find the distance
modulus-redshift relation in the case of metric (8) as follows:
\begin{equation}
m-M=-2.5\log (\frac \Lambda {12\pi })+2.5\log z+1.086(0.69+\frac z2-\frac{z^2%
}8+\frac{z^3}{24}-\ldots )
\end{equation}
A comparison (6) with (19) shows that (6) has a slope of 5 for small z while
the slope for (19) is 2.5 . This 
reveals that the FRW and static de Sitter frames both together do not belong to the same comoving frame of reference. The agreement of observational
data by (6) indicates that in the comoving frame of the FRW coordinate frame the galaxies are at rest. We may conclude that this test
confirms the FRW coordinates and discards the
static de Sitter coordinates as comoving. Thus in the
presence of a non-zero cosmological constant the de Sitter form (8) exhibits
problems and working in this coordinate as a comoving frame is physically unreasonable.

In fact according to this test, the validity of any solution of the field
equation which asymptotically approaches to the static de Sitter metric is under
question. The calculations for determining the line element for a point mass 
$m$ in the presence of $\Lambda $ are usually carried out in the second
coordinate frame i.e. asymptotically approach (8) as a boundary
condition. The Schwarzschild-de Sitter metric is  
\begin{equation}
ds^2=(1-\frac \Lambda 3\rho ^2-\frac{2m}\rho )dT^2-(1-\frac \Lambda 3\rho ^2-%
\frac{2m}\rho )^{-1}d\rho ^2-\rho ^2(d\theta ^2+\sin ^2\theta d\phi ^2)
\end{equation}
While this solution only maintains spherical symmetry, the physical subspace has homogeneity and isotropy symmetries which
shows a disadvantage of (20) in this respect. So if we are going to work in a comoving cosmological frame of reference, we need an alternative coordinate
frame in which the metric should 
asymptotically approach to the non-static de Sitter metric (1), and the isotropy and homogenity of the universe formulated by the cosmological principle is manifested in best way.

\noindent This has been done [8], and the result is

\begin{eqnarray}
ds^2 &=&\frac 12[\sqrt{(1-\frac{2M}\rho -\frac \Lambda 3\rho ^2)^2+\frac{%
4\Lambda }3\rho ^2}+(1-\frac{2M}\rho -\frac \Lambda 3\rho ^2)]dt^2  \nonumber
\\
&&-2e^{2\sqrt{\frac \Lambda 3}t}[[\sqrt{(1-\frac{2M}\rho -\frac \Lambda
3\rho ^2)^2+\frac{4\Lambda }3\rho ^2}+(1-\frac{2M}\rho -\frac \Lambda 3\rho
^2)]^{-1}dr^2  \nonumber \\
&&+\rho ^2(d\theta ^2+\sin ^2\theta \,d\varphi ^2)]  \nonumber \\
\rho  &\equiv &\,e^{(\sqrt{\frac \Lambda 3}t)}r \\
&&  \nonumber
\end{eqnarray}
On the basis of the above reasoning, eq. (20) is not a coordinate description of a cosmological comoving frame of reference. We should consider (21) as a solution for this system in comoving frame.

It is remarkable that metrics like (8) and (20) could be rejected as acceptable cosmological models according to Weyl's postulate which states that particles of the substratum lie in space-time on a congruence of timelike geodesics diverging from a point in the finite or infinite part. Evidently in (8) and (20) the world lines of comoving observers are not timelike geodesics in the whole space. Our conclusion may be considered as an observational confirmation of this postulate.

\end{document}